\documentclass[preprint, notoc]{JHEP3}
\pdfoutput=1
\usepackage{graphicx,subfig,amsmath,natbib}

\title{Constraining warm dark matter with cosmic shear power spectra}

\author{
Katarina Markovi\v{c}$^{a,b}$, Sarah Bridle$^{c}$, An\v{z}e Slosar$^{d}$, Jochen Weller$^{a,b,e}$\\
$^{a}$University Observatory Munich, Ludwig-Maximilian University, Scheinerstr. 1, 81679 Munich, Germany\\
$^{b}$Excellence Cluster Universe, Boltzmannstr. 2, 85748 Garching, Germany\\
$^{c}$Department of Physics and Astronomy, University College London, Gower Street, London WC1E 6BT, United Kingdom\\
$^{d}$Building 510A, Brookhaven National Laboratory, Upton, NY 11973-5000, USA\\
$^{e}$Max-Planck-Institut for Extraterrestrial Physics, Giessenbachstr., 85748 Garching, Germany\\
{\rm E-mail}: \email{markovic(at)usm.lmu.de, sarah.bridle(at)ucl.ac.uk, anze(at)bnl.gov, jochen.weller(at)usm.lmu.de}
}

\abstract{
We investigate potential constraints from cosmic shear on the dark matter particle mass, assuming all dark matter is made up of light thermal relic particles. Given the theoretical uncertainties involved in making cosmological predictions in such warm dark matter scenarios we use analytical fits to linear warm dark matter power spectra and compare 
(i) the halo model using a mass function evaluated from these linear power spectra and 
(ii) an analytical fit to the non-linear evolution of the linear power spectra.
We optimistically ignore the competing effect of baryons for this work. We find approach (ii) to be conservative compared to approach (i). We evaluate cosmological constraints using these methods, marginalising over four other cosmological parameters. Using the more conservative method we find that a Euclid-like weak lensing survey together with constraints from the Planck cosmic microwave background mission primary anisotropies could achieve a lower limit on the particle mass of 2.5 keV.
}

\preprint{arXiv:1009.0218}

\begin{document}

\section{Introduction}
\label{sec:intro}

In the second half of the 20th century, two competing theories for the growth of cosmological structure were proposed. In the cold dark matter (CDM) paradigm (\cite{1982ApJ...258..415P, 1984Natur.311..517B,1984ApJ...277..470P, 1985ApJ...292..371D}), dark matter possesses a negligible initial velocity dispersion and structure formation proceeds in a strictly hierarchical bottom-up manner: small haloes of dark matter form and later merge into larger ones, which in turn merge into even larger haloes. In these virialised dark matter structures  the baryons condense and form luminous objects in the Universe. In the hot dark matter (HDM) paradigm (\cite{1970A&A.....5...84Z,1980PhRvL..45.1980B,1983ApJ...274..443B,1988ApJ...333...24C}) on the other hand, the dark matter is ultra-relativistic in the very early stages of the Universe, erasing all structure on small scales. In these models, the most massive structures form first, producing ``Zeldovich pancakes'', that later produce smaller objects by fragmentation in a top-down manner.  An example of such an extremely energetic dark matter particle is a massive active neutrino.

By the end of the twentieth century it was clear that the hot dark matter paradigm cannot describe the measurements of the cosmic microwave background and the clustering of galaxies and that structure formation in the Universe is, at least overall, hierarchical (\cite{2010arXiv1001.4538K,2005MNRAS.362..505C, 2004PhRvD..69j3501T,2006JCAP...10..014S}).

However, there exist a number of observations that may indicate a lack of small scale perturbations in the initial density field, relative to that predicted in the standard $\Lambda$CDM paradigm. For example, it has long been known that CDM theory predicts many more small mass haloes than the number of dwarf galaxies that we see around the Milky Way (\cite{2007ApJ...657..262D} and references therein).  Similarly, cuspy galactic cores indicated in some observations are inconsistent with predictions of the CDM (\cite{1994Natur.370..629M,2005ApJ...621..757S}). Moreover, the angular momenta of dark matter haloes are considerably lower than those observed in spiral galaxies (\cite{2001ApJ...551..608S,2002MNRAS.336...55C,2008MNRAS.387..364Z}). There is also some discrepancy between the distribution of sizes of mini-voids in the local Universe and CDM predictions (\cite{2009MNRAS.399.1611T}).

These discrepancies might be resolved by accounting for certain astrophysical processes. Supernova feedback can extinguish star formation and further baryonic effects can also affect the properties of the dark matter density distribution in centres of haloes. However, a suppression of the primordial matter power spectrum on small scales is an attractive alternative. This is most easily achieved by giving dark matter some small initial velocity dispersion: not enough to break the very successful hierarchical structure formation, but enough to make a difference on small scales. Such models go under the name of warm dark matter (WDM) (\cite{Bode:2000gq,2001ApJ...559..516A}).

In warm dark matter models, dark matter particles free-streamed for a short period in the early Universe, before becoming non-relativistic. During this period, fluctuations on small scales were heavily suppressed and hence WDM models are characterised with a free-streaming wavenumber $k_{\rm fs}$, beyond which the fluctuations in the power spectrum of initial fluctuations are exponentially suppressed. This suppression is the main observational smoking gun of WDM models. For the models discussed in this paper, $\lambda_{\rm fs}\equiv 2 \pi / k_{\rm fs}$ is of the order of a Megaparsec.

Several microscopic models for warm dark matter have been proposed. The most common models contain sterile neutrinos (\cite{1994PhRvL..72...17D,2003PhRvD..68j3002F,2005PhLB..631..151A,Abazajian:2005xn,2006PhRvL..97z1302B,2008PhRvD..77f5014P,2008JCAP...06..031L,2009PhR...481....1K,Hamann:2010bk}) or gravitinos (\cite{1982PhRvL..48.1636B,1996PhLB..386..189B,2002PhLB..549..273F,2005PhRvL..95r1301C,Steffen:2006hw,2008PhLB..660..100T}) as dark matter particles.

In sterile neutrino models, the standard model of particle physics is extended with the addition of one or more singlet right-handed sterile neutrinos. The standard model active neutrinos then often receive the observed small masses through a see-saw mechanism. This implies that sterile neutrinos must be rather heavy, but the lightest of them naturally has a mass in the keV region, which makes it a suitable warm dark matter candidate. The simplest model of sterile neutrinos as WDM candidates assumes that these particles were produced at the same time as active neutrinos, but they never thermalised and were thus produced with a much reduced abundance due to their weak coupling. In the second class of models the gravitino appears as the supersymmetric partner of the graviton in supergravity models. If it has a mass in the keV range, it is a suitable WDM candidate. It belongs to a more general class of thermalised warm dark matter candidates. It is assumed that this class of particles achieved a full thermal equilibrium, but at an earlier stage, when the number of degrees of freedom was much higher and hence their relative temperature with respect to the CMB is much reduced. 

When describing WDM structure formation there are three relevant variables that describe a particular model of WDM: the temperature $T_{\rm WDM}$ of WDM particles, their mass $m_{\rm WDM}$ and their energy density $\omega_{\rm WDM}$. However, the equations of the evolution of perturbations can be re-parametrised as a two-parameter family of models, where one typical specifies $\omega_{\rm WDM}=\Omega_{\rm WDM} h^2$ (with $h=H_0/100$ ${\rm km}$ ${\rm s^{-1}Mpc^{-1}}$ denoting the reduced Hubble's constant and $\Omega_{\rm WDM}$ for the fractional energy density of WDM in units of critical density) and the ratio $m_{\rm WDM}/T_{\rm WDM}$. This allows one to make the following  one-to-one relation between the mass of the thermalised WDM particle $m_{\rm WDM}$, such as the gravitino, and the mass of the simplest sterile neutrino $m_{\rm \nu s}$, such that the two models have an identical impact on cosmology (\cite{Viel:2005qj})
\begin{equation}
  m_{\rm \nu s} = 4.43 \left(\frac{m_{\rm WDM}}{{\rm keV}}\right)^{4/3} \left(\frac{\omega_{\rm WDM}}{0.1225}\right)^{-1/3}{\rm keV} \hspace{1 cm}.\end{equation}
The difference comes from the fact that in the gravitino case the particle is fully thermalised,
the number of effective degrees of freedom being determined by mass and energy density of dark matter, while in the simplest sterile neutrino case the number of degrees of freedom is fixed, while abundance is determined by mass and energy density of dark matter. In the rest of this paper we work with fully thermalised WDM particles.

The above WDM models have been constrained using several observational techniques (e.g. \cite{Polisensky:2010rw,Boyanovsky:2007ay} etc.). The most popular is the observation of the flux power spectrum of the Lyman-$\alpha$ forest, which can indirectly measure the fluctuations in the dark matter density on scales between $\sim$~100~kpc and $\sim$~10~Mpc. The limits are $m_{\rm WDM}>4$~keV or equivalently $m_{\rm \nu s}>28$~keV at 95\% confidence limit (\cite{2008PhRvL.100d1304V}, see also \cite{Viel:2005qj,2006PhRvL..97s1303S}). For the simplest sterile neutrino model, these lower limits are at odds with the upper limits derived from X-ray observations, which come from the lack of observed diffuse X-ray background from sterile neutrino annihilation and set the limit $m_{\rm \nu s}<1.8$~keV at 95\% confidence limits (\cite{2006JETPL..83..133B}). We stress that these results rule out the simplest sterile neutrino models. However, there exist theoretical means of evading small-scale power constraints (see e.g. \cite{Boyarsky:2008mt} and references therein).

The distribution of matter can be observed using the bending of light by gravity, termed gravitational lensing. This causes a distortion in the apparent shapes of distant galaxies, which is described to first order by a shear. The effect is generally small but can be measured by statistically averaging many observations together, termed cosmic
shear. The two point correlation function of the shears of pairs of galaxies is a probe of the underlying matter power spectrum and thus of the nature of dark matter.

The latest cosmic shear measurements (\cite{Fu:2007qq}; \cite{Massey:2009fh}, \cite{Schrabback:2009ba}) have sufficient precision to just constrain one parameter for the amplitude and one for the shape of the matter power spectrum, usually written as the linear theory rms fluctuation in 8 $h^{-1}$ Mpc spheres, $\sigma_8$ and the matter density of the Universe, $\Omega_{\rm m}$. Cosmic shear has been identified as the method with the most potential to solve one of the major problems in cosmology: the nature of dark energy (\cite{Albrecht:2005p3300}; \cite{Peacock:2006kj}). There are therefore several upcoming large cosmic shear surveys both from the ground (Pan-STARRS\footnote{http://pan-starrs.ifa.hawaii.edu/}, the Dark Energy Survey\footnote{https://www.darkenergysurvey.org/}, HyperSuprimeCam\footnote{\cite{2006SPIE.6269E...9M}} and the Large Synoptic Survey Telescope\footnote{http://www.lsst.org/lsst/}) and space (Euclid\footnote{http://sci.esa.int/euclid} and WFIRST\footnote{http://wfirst.gsfc.nasa.gov/}), with additional surveys proposed.

Non-gravitational interaction between dark matter and baryonic matter is extremely weak or non-existent. Therefore, gravitational lensing is one of the most powerful probes of the nature of dark matter. Because it can detect the presence of dark matter directly, it provides some strong evidence for its existence (\cite{Clowe:2003tk}). Measuring gravitational lensing also enables us to map-out the 2 and even 3-dimensional distribution of dark matter (\cite{Kaiser:1992ps}, \cite{Hu:1999ek} and \cite{Taylor:2004ca}). We can also calculate the statistical properties of the matter distribution, which allow us to infer the matter density and the root-mean-squared amplitude of matter density fluctuations (\cite{Hoekstra:2002xs}). Since dark matter particle properties may directly influence the matter distribution, these measurements can additionally constrain the model for dark matter particle origin and nature (e.g. \cite{Schaefer:2008ku}), which is the topic of this paper.

In section two we describe how warm dark matter suppresses structure on small scales through free-streaming. We go on to discuss and compare two different methods of calculating the effects of non-linear evolution of the matter density field in the WDM case. We do not try to develop a new model for non-linear WDM structure as this is beyond the scope of this project. 
In the third section we calculate the weak lensing power spectrum for CDM and different models of WDM.
In section four we report our predictions for a Euclid-like survey and the potential of such a survey to constrain the WDM particle mass. We take a flat $\Lambda$CDM or $\Lambda$WDM Universe throughout. Our parameters have the values of WMAP7
(\cite{Larson:2010gs}):
the dimensionless Hubble expansion rate parameter, $h=0.71$, 
the present day baryon density in units of the critical density, $\Omega_{\rm b}=0.045$, 
the present day dark matter density in units of the critical density, $\Omega_{\rm m}=0.27$,
the cosmological constant energy density in units of the critical density, $\Omega_{\Lambda}=0.73$, 
the scalar primordial power spectrum slope, $n_{\rm s}=0.96$,
and the present day linear theory root-mean-square density fluctuation in 8 $h^{-1}$ Mpc spheres, $\sigma_8=0.80$.

\section{The effect of WDM on the distribution of matter}
\label{sec:wdmdom}

We begin by describing in more detail the evaporation of low mass halos through free-streaming and write down an approximate halo mass limit.
We then go on to how we implement the halo model of the Universe in WDM scenarios.
Finally we extrapolate the fitting formulae to CDM simulations in order to describe non-linear WDM structure.

\subsection{Free-streaming of WDM particles}
\label{sec:freestream}

Free-streaming results in suppressed formation of dark matter overdensities smaller than the free-streaming length. We define a ``free-streaming mass'', $M_{\rm fs}$ as the mass contained within a radius of $\lambda_{\rm fs}$ with the average density equal to the background matter density. In other words, it is the mass of a dark matter halo that formed from an initial linear overdensity of size equal to the distance of warm dark matter free-streaming,  $\lambda_{\rm fs}$. We calculate the free-streaming length as (Zentner and Bullock [2003]): 
\begin{equation}
\lambda_{\rm fs} \simeq 0.11 \left[\frac{\Omega_{\rm WDM}h^2}{0.15} \right]^{1/3}\left[\frac{m_{\rm WDM}}{\rm keV} \right]^{-4/3} {\rm Mpc}
\hspace{1 cm}
\end{equation}
and define a corresponding `free-streaming halo mass' as 
\begin{equation}
M_{\rm fs} = \frac{4\pi}{3} \left(\frac{\lambda_{\rm fs}}{2}\right)^{3}\rho_{\rm m}(z) \hspace{1cm}, 
\end{equation}
where $\rho_{m}(z)$ is the background matter density at redshift z.\\

\begin{figure}
\centering
\subfloat{\includegraphics[width=0.5\textwidth]{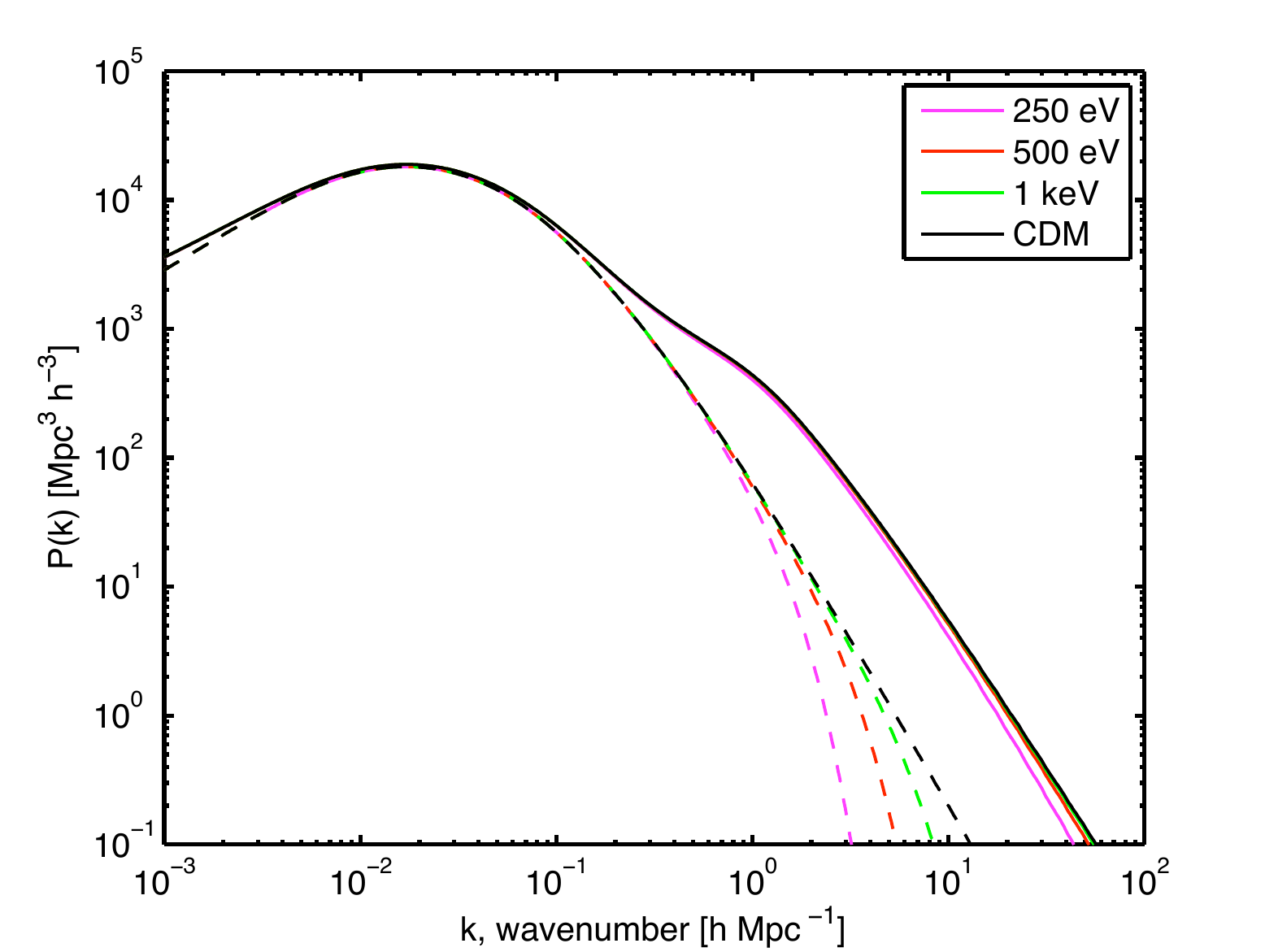}}
\subfloat{\includegraphics[width=0.5\textwidth]{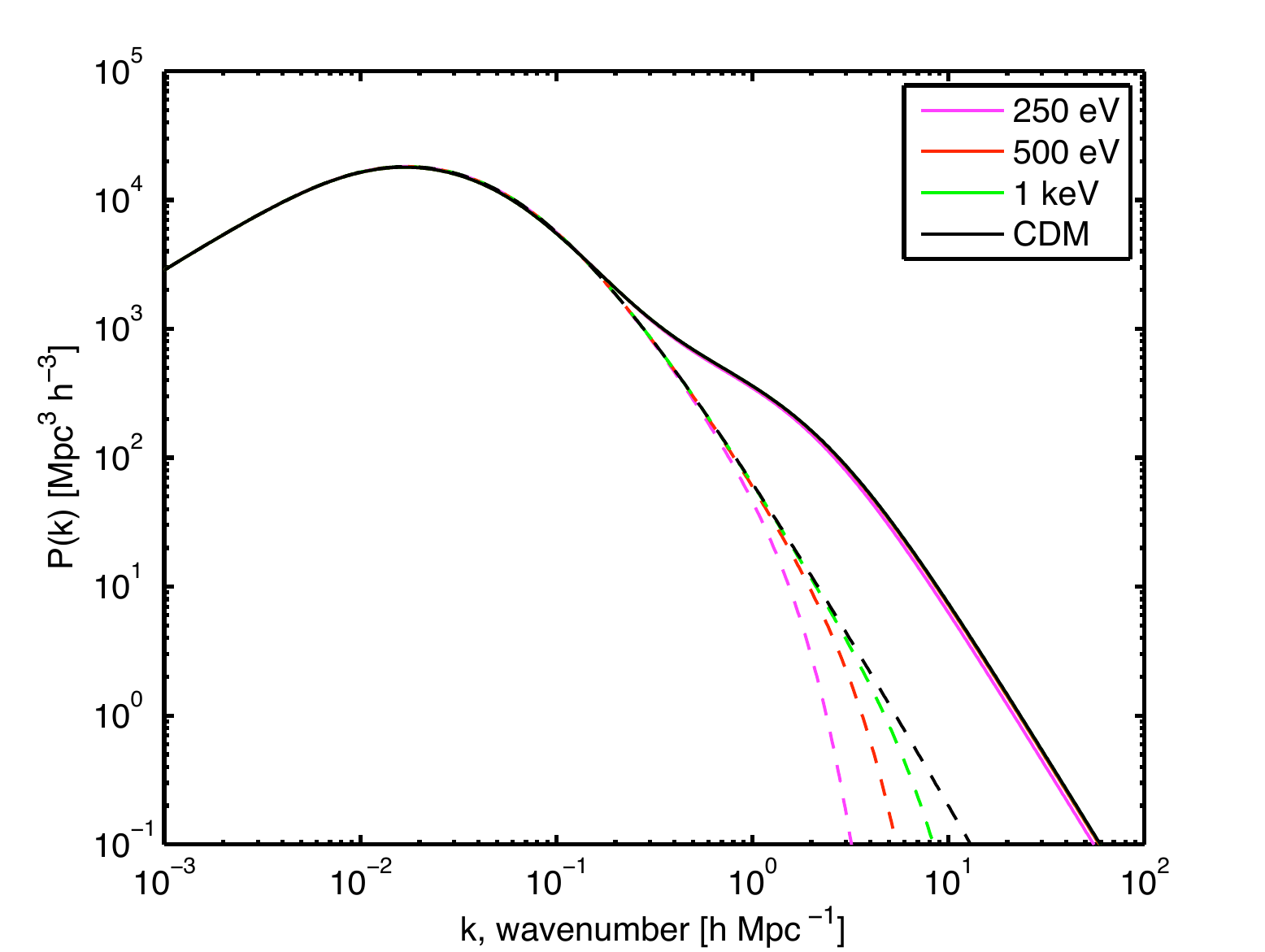}}
\caption{\small Matter power spectra today (zero redshift), suppressed by free-streaming of WDM particles. The dashed lines are the linear power spectra corresponding to, from left to right in each panel, 250 eV, 500 eV, 1 keV warm dark matter and cold dark matter. The dashed lines are the same in both plots. The solid lines are the corresponding non-linear power spectra. The non-linear power spectrum in the left panel uses the halo model and that in the right panel was calculated using the halofit formula. The slight excess of power on large scales in the halo model comes from neglecting the effect of `halo exclusion' (\cite{Smith:2010fh}).}
\label{fig:mpswdm}
\end{figure}

The power spectrum of matter distribution is modified by this erasure of small mass halos. We modify the CDM linear theory power spectrum shape to the WDM case through multiplication by a transfer function acting on the linear theory density power spectrum (see \cite{Boyanovsky:2008he} for a calculation of the transfer function for a general initial thermal distribution of DM particles - cold WIMP dark matter, thermal fermionic or bosonic dark matter):
\begin{equation}
T(k) \equiv \left(\frac{P_{\rm WDM}(k)}{P_{\rm CDM}(k)}\right)^{1/2}= [1 + (\alpha k)^{2\mu} ]^{-5/\mu}\hspace{1cm},
\end{equation}
(\cite{Bode:2000gq})
with adjustable parameter $\mu = 1.12$
(\cite{Viel:2005qj}) found by fitting to the result of Boltzman integration
(using CAMB by \cite{Lewis:1999bs} or CMBFAST by \cite{Seljak:1996is}). The break-scale parameter, $\alpha$ in the case of thermal relic DM is
\begin{equation}
\alpha = 0.049 \left(\frac{m_{\rm WDM}}{\rm keV}\right)^{-1.11}\left(\frac{\Omega_{\rm WDM}}{0.25}
\right)^{0.11}\left(\frac{h}{0.7}\right)^{1.22}h^{-1} \rm{Mpc}
\end{equation}
where $\Omega_{\rm WDM}$ is the WDM density parameter, which in this paper equals the total dark matter density parameter, $\Omega_{\rm DM}$, since all dark matter is assumed to be warm.
The dashed lines in Fig.~\ref{fig:mpswdm} show linear theory matter power spectra for a range of WDM masses. The lightest WDM particle mass shown (250 eV) causes the linear theory matter power spectrum to fall dramatically at a wavenumber significantly above 1 $h$ Mpc$^{-1}$. As also shown in \cite{Smith:2011pr}, the matter power spectrum of WDM starts to turn off well above the free-streaming scale, which changes the slope of the power spectrum to fall much more steeply than $n_{\rm eff}=\log P(k)/\log k=-3$, which is the slope for standard, bottom-up structure formation (\cite{White:2000sy}, \cite{Knebe:2003hs}). When, at higher k, the slope falls more steeply than that, bottom-up structure formation halts and top-down fragmentation might occur. This effect significantly alters the mass functions and since the effect starts at scales larger than the free-streaming scale, this influences halo masses above the `free-streaming' mass (see Fig.~\ref{fig:massfnandratio}).

\subsection{The halo model approach}
\label{sec:halo}

\begin{figure}
  \centering
\includegraphics[width=0.7\textwidth]{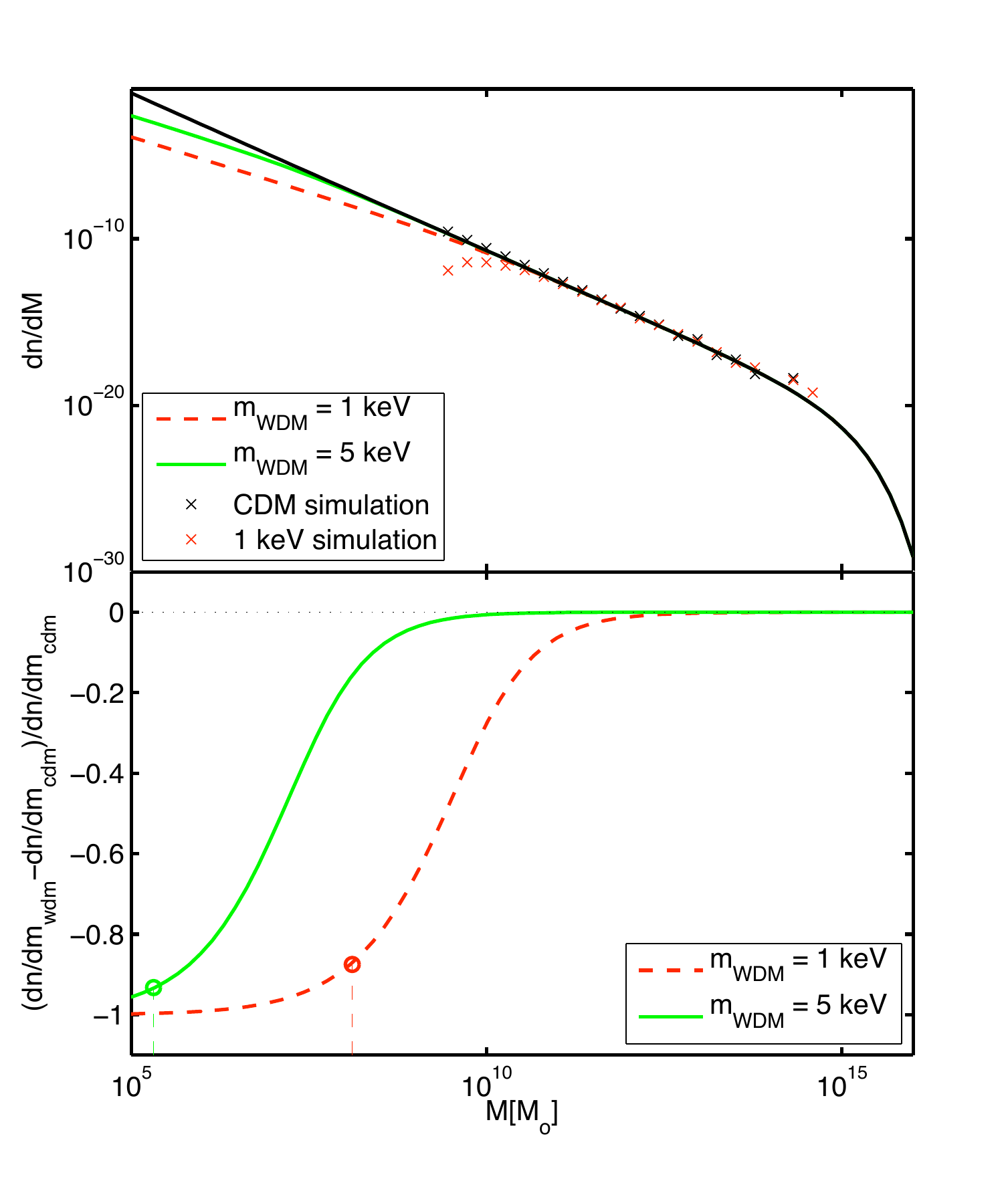}
\caption{\small The top figure shows Sheth-Tormen mass functions suppressed by free-streaming of WDM particles for two different warm dark matter scenarios: 1 keV particles (dashed red line), 5 keV particles (solid green line) and CDM (solid black line). The bottom figure shows the fractional difference between the WDM and CDM mass functions for the 2 different warm scenarios. The circles mark the halo mass on the x-axis that corresponds to two-times the free-streaming radius of each WDM particle (or eight-times the ``free-streaming mass''). Data points from the simulation by \cite{Zavala:2009ms} are marked with crosses on the upper panel: the red crosses correspond to the same WDM scenario as the red dashed line (only data points above the limiting mass of the simulation are shown).} 
\label{fig:massfnandratio}
\end{figure}
The halo model assumes the Universe is made up of halos with positions sampled from the linear theory matter distribution. As a result, there are two main contributions to the non-linear matter power spectrum (\cite{Ma:2000ik}, \cite{Seljak:2000gq}, \cite{Peacock:2000qk}). Firstly, the two-halo term, which dominates on large scales, encodes the correlation between different haloes and is equal to the linear matter power spectrum on large scales. Secondly, the one-halo term refers to the correlations within a halo and therefore depends on the density profile of the halo. Both terms depend on the number of halos as a function of halo mass, which can be found to a reasonable approximation using analytic arguments (Press \& Schechter 1975) or more usually measured from numerical simulations and modified according to the underlying matter power spectrum (e.g.  \cite{Bertschinger:1900nj}, \cite{Sheth:1999mn} and in the context of weak lensing, \cite{Cooray:2000ry} and \cite{Takada:2002qq}).
For reasons of simplicity we assume spherical collapse in our halo model calculations.

We extend the halo model to WDM scenarios by modifying the ingredients. We use the WDM linear power spectrum described in the previous subsection to calculate a new mass function using the \cite{Sheth:1999mn} prescription. We make the conservative assumption that the halo profiles are unchanged relative to CDM. It is in the one-halo term of the power spectrum that the effects of free-streaming of WDM are seen most strongly. This is because of the difference in the root-mean-square fluctuation, $\sigma(R)$, which becomes suppressed at small $R$ in a WDM Universe.

For a detailed description of these calculations, see Appendix \ref{sec:appA}. We choose not to normalise the mass functions because the result of free-streaming is the suppression of the formation of small structures, not by accreting the excess matter onto larger haloes, but rather by resulting in a larger uncollapsed background matter density. In contrast, we do normalise the halo bias by requiring
\begin{equation} \label{eq:normalisation}
\frac{1}{\rho_{\rm m,0}}\int b(M)\frac{dn}{dM} M dM = 1 \hspace{1cm}.
\end{equation}
This ensures that our model reproduces the linear predictions in the limit of $k\rightarrow 0$.

We explore the effect of WDM on the mass functions in Fig.~\ref{fig:massfnandratio}. As expected, the number density for the smallest haloes is reduced in the case of WDM. This is shown most visibly in the lower panel of  Fig.~\ref{fig:massfnandratio} which shows the suppression on a linear scale. This is useful for comparison to the general assumption of the absence of haloes below the ``free-streaming'' mass.

The points in the upper panel of Fig.~\ref{fig:massfnandratio} make a comparison to an N-body simulation by \cite{Zavala:2009ms}. This indicates that the Sheth-Tormen mass functions underestimate the strength of this suppression. Indeed, comparing the simulated mass function for the 1 keV particle to the Sheth-Tormen version, we find a very large discrepancy around the free-streaming mass. The simulated mass function declines to zero much more steeply than the Sheth-Tormen. This suggests that we underestimate the effect of WDM on the mass functions. New models need to be developed to calculate the WDM mass functions correctly, most likely on the basis of further WDM simulations. We have removed the first three points in the simulated mass function for the lowest three halo mass bins. This is because these are too affected by Poisson noise (\cite{Zavala:2009ms}).

\subsection{Non-linear theory fitting formula (halofit)}
\label{sec:halofit}

In the $\Lambda$CDM model, the non-linear evolution of the matter power spectrum can be calculated from the primordial linear matter power spectrum, using the halofit formula of \cite{Smith:2002dz}. This method is a fitting function based on the halo model, but very accurately reproduces the results of many CDM numerical simulations. It is more accurate and more general than the fitting function from \cite{Peacock:1996ci}. In this paper we make the large extrapolation of assuming that the halofit method is valid for WDM. In Fig.~\ref{fig:mpswdm} we compare non-linear matter power spectra calculated using this method to power spectra calculated using the halo model formalism described in the previous section.

Non-linear evolution of the matter power spectrum washes out the difference between WDM and CDM on small scales, which is apparent when comparing the linear spectra to the non-linear spectra in either panel of Fig.~\ref{fig:mpswdm}. This is due to the collapse of structures, which significantly boosts the power on small scales and more than compensates for any missing structure (\cite{Boehm:2005p2215}). The information contained in the matter power spectrum on cosmological parameters is encoded in the shape of the matter power spectrum. If the matter power spectrum is modified to smooth out the shape (e.g. by non-linear growth) then some of the information contained in the shape will be no longer be available, just as for any smoothing of an observable. In particular on small scales there is very little power to begin with, so when power cascades down from larger scales the little power is swamped. For any mildly noisy observation the information originally encoded in the small scales would be much harder to recover after the power had cascaded down, i.e. information would have been lost.

\section{Shear power spectra}
\label{sec:wlps}

\begin{figure}
 \centering
\includegraphics[width=0.7\textwidth]{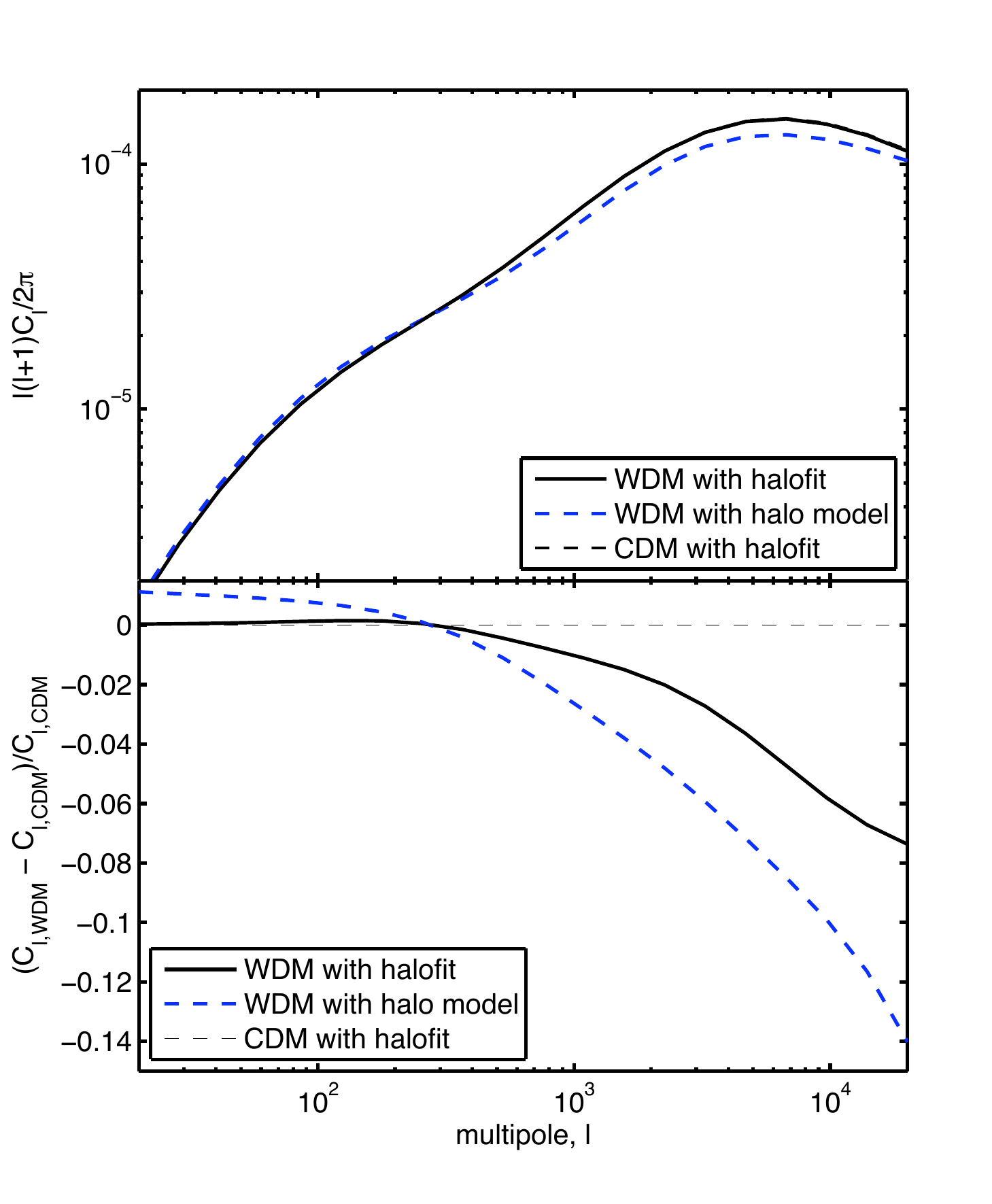}
\caption{\small
In the upper panel we plot the weak lensing auto-correlation power spectra for tomographic redshift bin 5, for 1 keV WDM particles and CDM.
The lower panel shows the fractional differences between 1 keV WDM and CDM spectra for both methods: 
the thick solid (black) and the thick dashed (blue) lines were found using the WDM linear theory matter power spectrum with the non-linear contribution calculated with the halofit method and the halo model respectively.
The thin dashed line was found using halofit with CDM in both panels. However it cannot be seen in the panel above, because it is only different from the thick (black) line by 1\%. 
Note that the excess of power in the halo model on large scales is due to the normalisation of the halo bias and is unphysical.} \label{fig:wlpsmethods}
\end{figure}
\begin{figure}
 \centering
\includegraphics[width=0.7\textwidth]{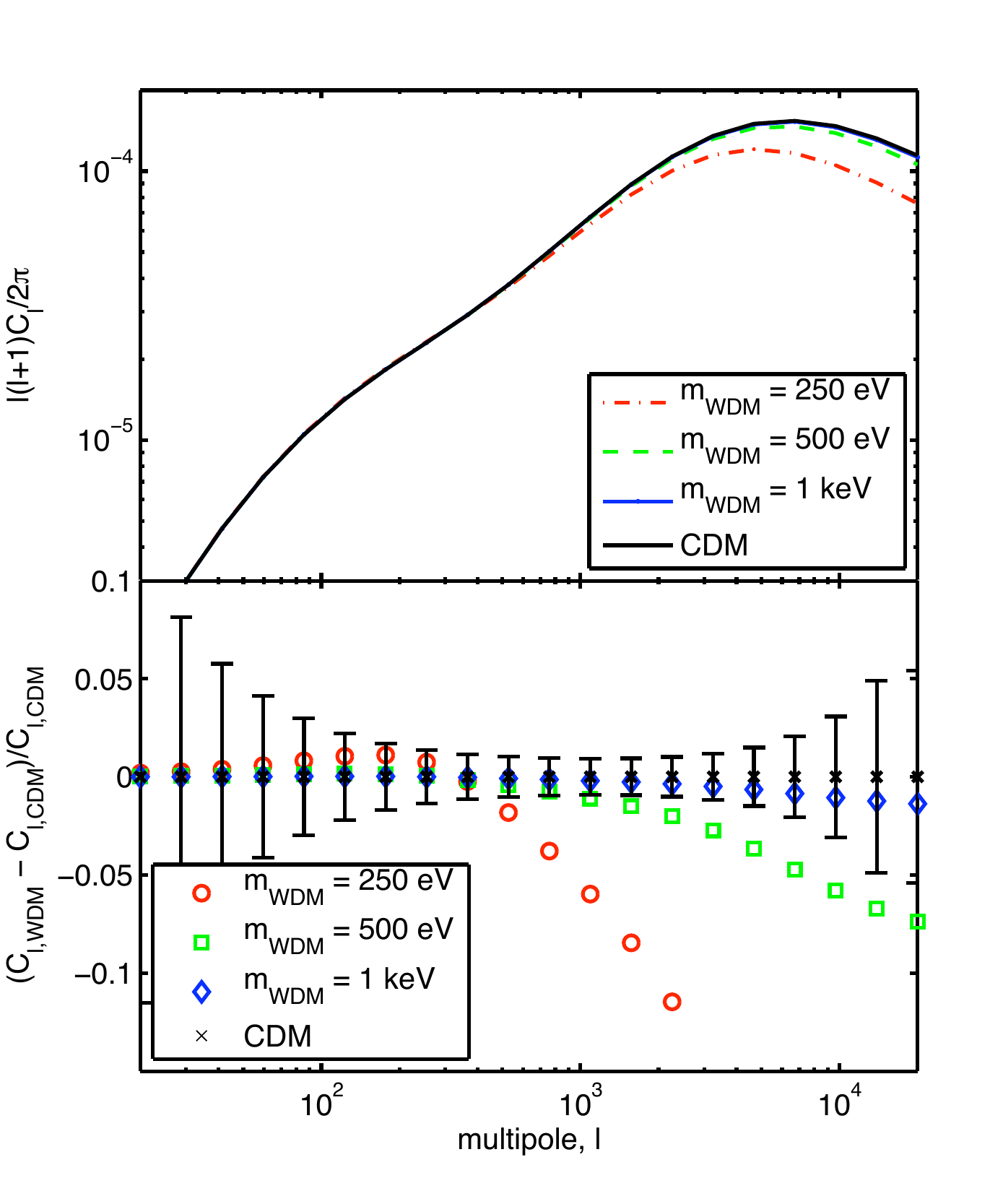}
\caption{\small 
In the upper panel we plot the weak lensing auto-correlation power spectrum for tomographic redshift bin 5. We show the power spectra for CDM (solid black) and three different WDM particle masses:
250 eV, 500 eV and 1 keV denoted with the dashed (red), dot-dashed (green) and solid (blue) lines respectively. The effect of the 1 keV particle is not visible on this plot and the blue line overlaps with the black. %\\
In the lower panel we plot fractional differences between the WDM and CDM power spectra shown in the upper panel. The (red) circles, the (green) squares and the (blue) diamonds represent the 250 eV, 500 eV and 1 keV particles respectively. Furthermore, here we plot the error bars for a Euclid-like survey in 20 multipole bins.}  
\label{fig:wlpswdm}
\end{figure}
In this paper we consider how to measure the WDM particle mass using observations of cosmic shear power spectra. From an observer's point of view, the image of each galaxy in the Universe is distorted by gravitational lensing effects of all intervening matter. Therefore the cosmic shear power spectra are closely related to the matter power spectrum integrated over redshift. Future surveys are expected to use broadband photometry to estimate the redshifts of the observed galaxies. This should allow shear power spectra to be calculated at a range of different redshifts, and also allow cross power spectra between redshifts (see \cite{Csabai:2002yk} for a review). We first calculate and compare cosmic shear power spectra calculated using the methods described in the previous section. We plot the comparison in Fig. \ref{fig:wlpsmethods}. We then show power spectra for different WDM masses and compare visually with the uncertainties from an upcoming survey in Fig.~\ref{fig:wlpswdm}.

We consider a cosmic shear survey which has a number of galaxies per unit redshift
\begin{equation} \label{eq:nofz}
n(z)=z^\alpha e^{-(z/z_0)^\beta} \hspace{1cm} ,
\end{equation}
(\cite{Smail:1994sx}) where $\alpha = 2$, $\beta = 1.5$ and $z_0 = z_{\rm m}/1.412$, where $z_{\rm m} = 0.9$ is the median redshift of the survey. For example, this is reasonable for a Euclid-like survey (\cite{Amara:2006kp}). We assume a photometric redshift uncertainty of  $\delta (1+z) = 0.05(1+z)$ and take top-hat photometric redshift bins containing equal numbers of galaxies (as in \cite{Amara:2006kp}). 
We use 35 galaxies per square arcminute and take $f_{\rm sky}$, the fraction of the sky covered by the survey, to be a half. %K
We assume there are no catastrophic outliers. We calculate power spectra at 20 spherical harmonic multipoles in the range $20<\ell<20000$ (roughly $0.0065<k<6.5$ at redshift 0.9, or down to scales of order 1 arcminute). 

The lensing power spectra are given by
\begin{equation}\label{eq:lensing}
C_{ij}(l) = \int_{0}^{\chi_{\rm H}} d\chi_{\rm l} W_{i}(\chi_{\rm l})W_{j}(\chi_{\rm l})\chi_{\rm l}^{-2}P_{\rm nl}\left(k=\frac{l}{\chi_{\rm l}},\chi_{\rm l}\right) \hspace{1cm} ,
\end{equation}
where $\chi_{\rm l}(z_{\rm l})$ is the comoving distance to the lens at redshift $z_{\rm l}$ and $W_{i}$ is the lensing weight in the tomographic bin {\it i}:
\begin{equation}
W_{i}(z_{\rm l}) = \rho_{\rm m,0} \int_{z_{\rm l}}^{z_{\rm max}} \left[\frac{n_{i}(z_{\rm s})}{\Sigma_{\rm crit}(z_{\rm l},z_{\rm s})}\right] dz_{\rm s} \hspace{1cm} ,
\end{equation}
where
\begin{equation}\label{eq:sigmacrit}
\Sigma_{\rm crit}(z_{\rm l},z_{\rm s}) = \frac{c^2}{4\pi G}\frac{\chi_{\rm s}}{\chi_{\rm ls}\chi_{\rm l}}\frac{1}{(1+z_{\rm l})} \hspace{1cm},
\end{equation}
and the subscripts s, l and ls denote the distance to the source, the distance to the lens and the distance between the lens and source respectively.

In the halo model one can simplify the above equation using some approximations (e.g. \cite{Cooray:2000zy}) in the halo model non-linear matter power spectrum so that the lensing power spectra are made up of one-halo and two-halo terms, 
\begin{equation} \label{halowlps}
C_{ij}(l) = C_{ij}^{\rm 1h}(l) + C_{ij}^{\rm 2h}(l) \hspace{1cm} .
\end{equation}
which are given by
\begin{eqnarray}
C_{ij}^{\rm 1h}(l) &=& \int_0^{z_{\rm max}} dz \frac{d^2V}{dzd\Omega}
\int_{M_{\rm min}}^{M_{\rm max}} dM
\frac{dn(M,z)}{dM} \tilde{\kappa}_{i}(l,M,z)\tilde{\kappa}_j(l,M,z)
\label{eq:1h}\\
C_{ij}^{\rm 2h}(l) &=& \int_0^{z_{\rm max}} dz \frac{d^2V}{dzd\Omega}P\left(\frac{l}{\chi},z\right) T_{i}(l,z)T_{j}(l,z) \hspace{0.5cm},\\
\label{eq:Ti}
T_{i}(l,z)
&=&\int_{M_{\rm min}}^{M_{\rm max}} \frac{dn(M,z)}{dM}b(M,z)
\tilde{\kappa}_{i}(l,M,z) dM
\end{eqnarray}
(\cite{Cooray:2000ry}):
where $\tilde{\kappa}_{i}(l,M,z)$ is the 2-dimensional Fourier transform of the lensing convergence for which one needs to integrate the NFW halo density profile (\cite{Navarro:1996gj}), truncating the integral at $R_{180}$, the spherical radius, where the density contrast, $\Delta = 180$. We use the analytic results given in \cite{Takada:2002qq}. The lensing convergence signal in redshift bin $i$ is found by the following equation:
\begin{equation}
\kappa_{i}(\theta,M,z_{\rm l})=\int_{z_{\rm l}}^{z_{\rm max}} n_{i}(z_{\rm s})\kappa(\theta,M,z_{\rm l},z_{\rm s}) dz_{\rm s}\hspace{1cm}. \label{eq:kappa_i}
\end{equation}

We use the WDM linear theory matter power spectrum within the full halo model analysis,
and calculate the resulting shear power spectrum, shown by the dashed line in Fig.~\ref{fig:wlpsmethods} for the case of a 1 keV WDM particle mass, for the autocorrelation of the fifth redshift bin with itself. As expected, the cosmic shear power spectrum amplitude is reduced on medium to small scales due to the lack of contributions to the one-halo term from smaller halos. On mid-range scales the two-halo term is very slightly greater than the CDM power spectrum due to the renormalisation of the halo bias given in Eq.~\ref{eq:normalisation}. The mass function is decreased on small scales by WDM and therefore the bias must increase on all scales to satisfy the bias normalisation constraint. For the matter power spectrum this exactly preserves the large scale power, but the lensing two-halo term weights the halos by their lensing effect ($\tilde{\kappa}$ in Eq.~\ref{eq:Ti}), which is largest for high mass halos, where the mass function was unchanged but the bias is larger. Consequently, as expected, the lensing power spectrum is increased on larger scales.

We then use the linear theory WDM function within halofit to calculate the non-linear matter power spectrum, which we then insert into the standard cosmic shear power spectrum integrals. This is shown by the black solid line in Fig.~\ref{fig:wlpsmethods}. The effect is similar to that from the halo method but the deviation from CDM is about half the size.

In using these two different methods we have attempted to somewhat span the range of possibilities given the present theoretical uncertainties. They all agree surprisingly well, to within a factor of two. To be conservative, we use the halofit method from now on, unless otherwise stated.

For illustration we show the halofit results for three different WDM particle masses in Fig.~\ref{fig:wlpswdm}. As expected, the suppression of the shear power spectrum is strongest and reaches the largest scales for the lowest DM particle masses. This reflects the scarcity of low mass lenses in the WDM Universe. The effect of WDM on the shear power spectrum (figure \ref{fig:wlpswdm}) is more prominent than that on the non-linear matter power spectrum at redshift zero (figure \ref{fig:mpswdm}). WDM has a greater effect on the linear theory matter power spectrum than the non-linear matter power spectrum and therefore has a greater effect on the non-linear matter power spectrum at higher redshifts, where there is less non-linearity. Since the shear power spectrum is an integral of the matter power spectrum over redshift, WDM has a greater effect on the shear power spectrum than on the redshift zero non-linear matter power spectrum. We also see that the effect of halving the WDM particle mass is to more than double the perturbation to the CDM power spectrum i.e. the change in the deviation from CDM is non-linear in the inverse particle mass. We return to this in the following section.

We compare the lensing power spectra in Fig.~\ref{fig:wlpswdm} to predicted statistical uncertainties on 20 logarithmically spaced bins in $\ell$ for this redshift bin. Here and in the following section, we assume a Euclid-like survey with 35 galaxies per square arcminute covering half the sky with shear measurement error on each galaxy of  $\sigma_\gamma = 0.35 / \sqrt{2}$. (\cite{Refregier:2010ss}). The comparison is only indicative, since there are strong correlations between $\ell$ bins on small scales and between the 90 different redshift cross power-spectra (not shown). However, we already see that very high WDM particle masses around 250 eV will be easily ruled out.

\section{Forecasting constraints on the WDM particle mass}
\label{sec:results}

\begin{figure}
\centering
\includegraphics[width=\textwidth]{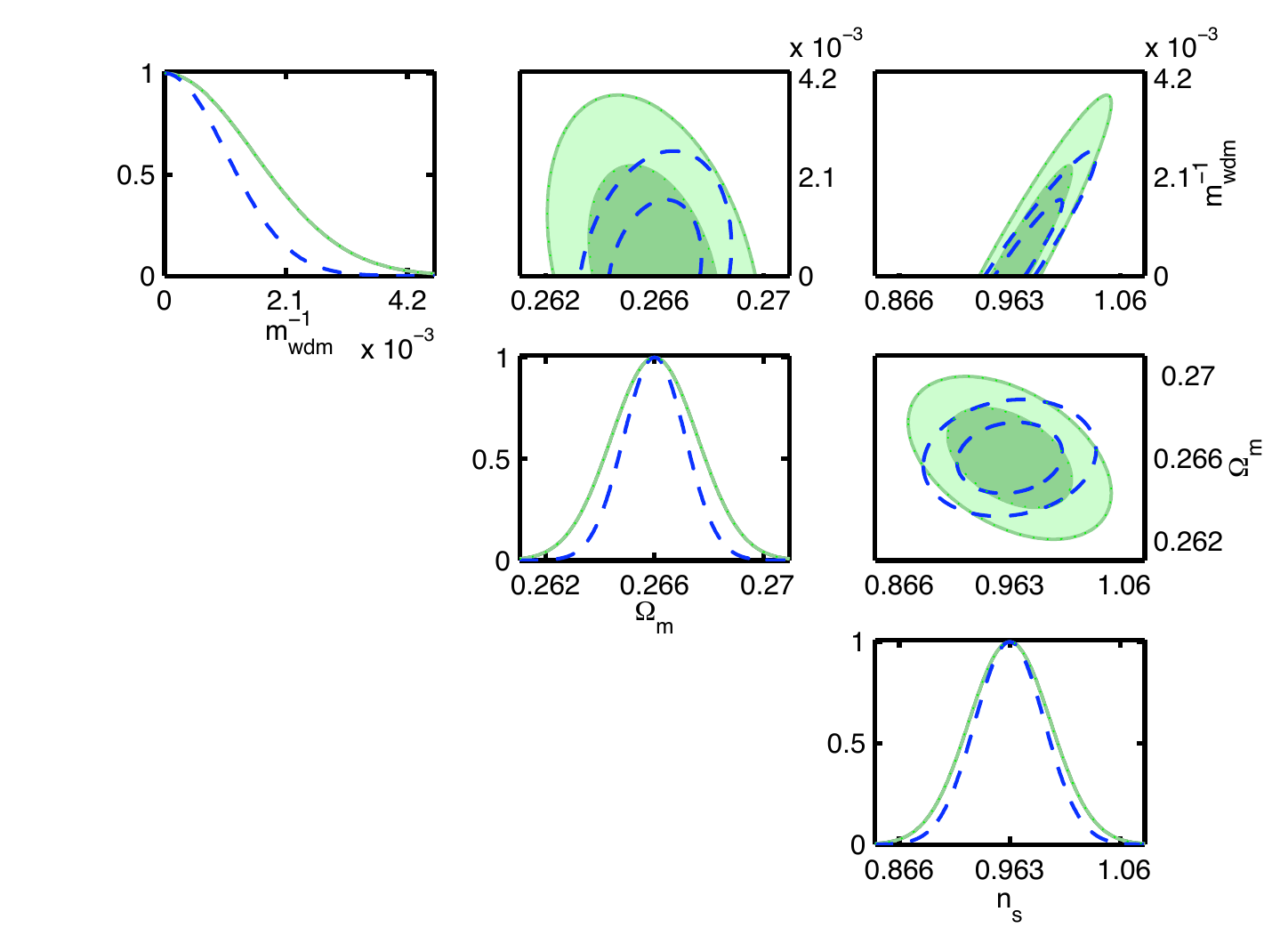}
\caption{\small 68 and 95 \% confidence likelihood contours for both the halo model (blue dashed) and the halofit (green solid) approaches, marginalised over the power spectrum amplitude parameter $A_{\rm s}$ and shape parameter $\Gamma$. The units on $m_{\rm WDM}^{-1}$ are eV$^{-1}$.}
\label{fig:ellipses}
\end{figure}
\begin{figure}
\centering
\includegraphics[width=\textwidth]{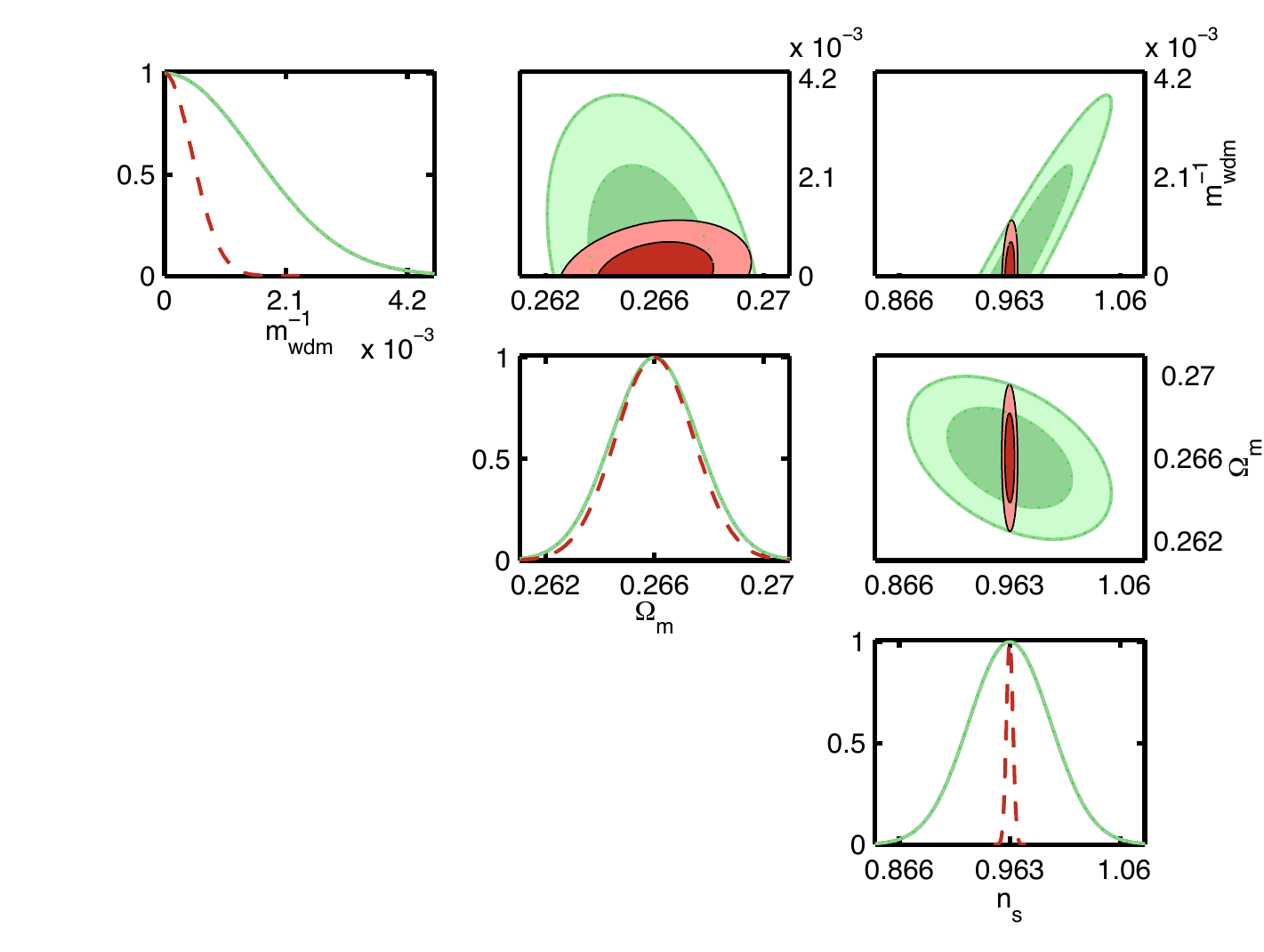}
\caption{\small
68 and 95 \% confidence likelihood contours for the halofit approach using cosmic shear (thick green solid) and cosmic shear combined with Planck forecasts (red dashed line and thin black solid contours shaded red). The units on $m_{\rm WDM}^{-1}$ are eV$^{-1}$.} 
\label{fig:ellipses_CMB}
\end{figure}
In this section we forecast constraints on the WDM particle mass using a Fisher matrix (FM) analysis. We vary four other cosmological parameters in the cosmic shear analysis and compare constraints between the halofit and halo model approaches. We then combine the halofit results with forecast constraints from the Cosmic Microwave Background (CMB).

We use CDM as our fiducial model and place limits on the inverse WDM particle mass so that our fidicual model has a parameter value of zero, corresponding to an effectively infinite WDM particle mass (in practice we use $m_{\rm WDM} = 100$ MeV). We vary the following additional parameters: the total matter density today, $\Omega_{\rm m}$, the primordial perturbation spectral index, $n_{\rm s}$, the primordial spectral amplitude, $A_{\rm s}$ and the matter power spectrum shape parameter, $\Gamma=e^{-2\Omega_{\rm b}h}\Omega_{\rm m}h$. We use $A_{\rm s}$ instead of $\sigma_8$ to normalise the matter power spectrum for convenience when combining with the CMB constraints. For the CDM linear theory matter power spectrum we use a fitting formula for the matter power spectrum from \cite{Ma:1996za} which uses the shape parameter $\Gamma$. Cosmic shear alone is very insensitive to $\Omega_{\rm b}$
and $h$ individually.

The Fisher matrix for cosmic shear is given as:
\begin{equation}
F_{\alpha,\beta} = \sum_{l}\sum_{(i,j),(m,n)} \frac{\partial C_{ij}(l)}{\partial \phi_{\alpha}} \left[{\rm Cov}[C_{ij}(l),C_{mn}(l)] \right]^{-1} \frac{\partial C_{mn}(l)}{\partial \phi_{\beta}}  \hspace{1cm},
\end{equation}
where $C_{ij}$ is the shear power spectrum of bin $i$ correlated with bin $j$, $\phi_{\alpha}$ are the parameters over which we marginalise and Cov$(C_{ij}(l),C_{mn}(l))$ is the covariance matrix of bins $i$, $j$, $m$ and $n$. We use equation (14) in \cite{Takada:2003ef} to find the elements of the covariance matrix, after adding the shot noise term coming from the shear error, $\sigma_{\gamma}$:
\begin{equation}
\bar{C}_{ij}(l) = C_{ij}(l)+\delta_{ij}\frac{\sigma^{2}_{\gamma}}{\bar{n}_{i}} \hspace{1cm},
\end{equation}
where $\bar{n}_{i}$ is the average number density of galaxies in redshift bin $i$.

The FM analysis assumes that the first non-vanishing order of the Tayor expansion of the log-likelihood is quadratic. However for CDM as the fiducial model, this is not the case for $m_{\rm WDM}^{-1}$ (or $m_{\rm WDM}$) as a parameter. Therefore the standard FM approach breaks down. As a result, the error found from FM analysis depends on the fiducial model chosen, errors are asymmetric around the fiducial value, and it turns out that the power spectra are completely insensitive to the inverse WDM particle mass at small values of $m_{\rm WDM}^{-1}$. Therefore it is not possible to use a very small step size in this parameter when calculating the derivative of the power spectrum in the Fisher matrix analysis. We discuss how we deal with this in Appendix~\ref{appFish}. Briefly, we tune the step size in the inverse WDM particle mass to equal the 1-sigma error bar on this parameter.

We show the resulting parameter constraints on $m_{\rm WDM}$, $\Omega_{\rm m}$ and $n_{\rm s}$ in Fig.~\ref{fig:ellipses}, marginalised over the power spectrum amplitude and shape. The halofit results are shown by the solid contours and the halo model by the dashed curves. The agreement between the uncertainties on each parameter is good, with the halo model giving slightly tighter constraints, as expected from the lensing power spectra shown in Fig.~\ref{fig:wlpsmethods}. The degeneracy direction between $m^{-1}_{\rm WDM}$ and $\Omega_{\rm m}$ is slightly different between the two methods. We have verified that on fixing all other cosmological parameters the degeneracy direction is extremely similar for the two methods, so the difference arises from the interplay between changing the WDM mass simultaneously with other parameters including the large scale amplitude parameter $A_{\rm s}$. The degeneracy direction between $m^{-1}_{\rm WDM}$ and $n_{\rm s}$ is as expected: increasing $m^{-1}_{\rm WDM}$ causes a decrease in power on small scales, which can be partially compensated by increasing the power spectrum index $n_{\rm s}$.

The marginalised errors that result from this analysis can be found in Table \ref{tab:errs}, using a one-tailed confidence limit for $m^{-1}_{\rm WDM}$. We find approximately 30\% smaller errors when using the halo model, compared to the halofit method.

We include in the analysis the Fisher matrix for the Planck CMB experiment in addition to the weak lensing Fisher matrix presented above.  We derive the cosmological constraints from Planck following the description by the DETF \cite{Albrecht:2006um} (for a more detailed discussion see \cite{Rassat:2008ja}). We assume that we will only use the 143 GHz channel data. This channel has a beam of $\theta_{\rm fwhm} = 7.1'$ and sensitivities in temperature of T = 2.2 $\mu$K/K and in polarization of P = 4.2 $\mu$K/K. We take $f_{\rm sky} = 0.80$ as the sky fraction in order to account for galactic foregrounds. We use as a minimum $l$, $l_{\rm min} = 30$ in order to avoid problems with polarization foregrounds. As described in the DETF report (\cite{Albrecht:2006um}) we choose as fiducial parameter set: $\omega_{\rm m}$, $\theta_{\rm s}$ , $\ln$$A_{\rm s}$, $\omega_{\rm b}$, $n_{\rm s}$, $\tau$, where $\theta_{\rm s}$ is the angular size of the sound horizon at last scattering, $\ln$$A_{\rm s}$ is the logarithm of the primordial amplitude of scalar perturbations and $\tau$ is the optical depth due to reionization. After marginalization over the optical depth and the physical baryon density, $\omega_{b}$ we then calculate the Planck CMB Fisher matrix in the parameters: $\Gamma$, $A_{\rm s}$, $n_{\rm s}$, $\Omega_{\rm m}$  by using the appropriate Jacobian of the involved parameter transformation (\cite{Rassat:2008ja} or \cite{Eisenstein:1998tu}). We point out that the Planck FM is computed {\em without} including the WDM paradigm. This is based on the assumption that the scales affected by WDM with $m_{\rm WDM}\gtrsim100$ eV are far away from the scales probed by CMB experiments. Therefore, while it adds quite strong constraints on the other cosmological parameters, especially on the curvature, the Planck FM does not add any constraints on $m_{\rm WDM}^{-1}$.

The parameter constraints are shown in Fig.~\ref{fig:ellipses_CMB} for the halofit method. We see that the very tight constraint on $n_{\rm s}$ from the CMB breaks the degeneracy between $n_{\rm s}$ and $m_{\rm WDM}$ from cosmic shear alone, to produce a tighter constraint on  $m_{\rm WDM}$. The single parameter marginalised results for halofit and the halo model methods are shown in Table~\ref{tab:errs}.

\begin{table}
\begin{center}
\begin{tabular}{| c || c || c | c | c |}
	\hline
	& \small Fiducial value & \small {\it Halo model} & \multicolumn{2}{|c|}{\small {\it Halofit}} \\
	& & \small Shear only & \small Shear only & \small Shear + Planck \\
	\hline \hline
	$m_{\rm WDM}$ & CDM & $>$ 935 eV & $>$ 645 eV & $>$ 2500 eV \\
	$m_{\rm WDM}^{-1}$ $\left[eV^{-1}\right]$ \rule{0cm}{0.5cm} & $10^{-8}$($\approx0$) & +0.00107 & +0.00155 & +0.00040 \\ \hline
	$\Omega_{\rm m}$ & 0.27 & $\pm$0.0011 & $\pm$0.0016 & $\pm$0.0012 \\ \hline
	$n_{\rm s}$ & 0.96 & $\pm$0.031 & $\pm$0.035 & $\pm$0.002 \\ \hline
	$A_{\rm s}$ [$h^{-3}Mpc^{3}$] & $1.3\times10^5$ & $\pm0.119\times10^5$ & $\pm0.183\times10^5$ & $\pm0.005\times10^5$ \\ \hline
	$\Gamma$ & 0.18 & $\pm$0.0090 & $\pm$0.0123 &  $\pm$0.0005 \\ \hline
\end{tabular}
\caption{\small Table of 68 per cent confidence limits for a flat $\Lambda$WDM cosmology marginalised over 4 other parameters in each case. The errors on the inverse particle mass, $m_{\rm WDM}^{-1}$, are single tailed limits.}
\label{tab:errs}
\end{center}
\end{table}

\section{Discussion}
\label{sec:disc}

We have made the first estimate of constraints on the WDM particle mass from cosmic shear power spectra. We compared approaches to estimating the impact of warm dark matter on cosmic lensing. We found the halofit approach of \cite{Smith:2002dz} more conservative that the halo model. We comment on the difference below. We find also that the effects on the mass functions (see Fig. \ref{fig:massfnandratio}) reach on mass scales significantly larger than the ``free-streaming'' mass. We made forecasts for parameter constraints from a Euclid-like cosmic shear survey combined with Planck and found that if the true cosmological model were CDM, we could place a limit of $m_{\rm WDM}>2.5$ keV at 68 per cent confidence.

Our calculations are conservative in several ways. Firstly, we use a transfer function to relate the CDM linear theory matter power spectrum to the WDM linear theory power spectrum, from which we calculate the mass function. We showed a comparison with a mass function from simulations which suggest a much bigger supression of the number of lower mass halos. To improve our constraints it would be necessary to have a fitting formula to simulations giving the mass function as a function of WDM particle mass, extending to smaller masses than the present simulations. Ideally, a better physical model for the non-linear collapse of WDM structures would be developed. In addition, the new WDM models for the mass functions should be physical enough to translate also into the halofit method and explain the decrease in signal due to WDM in the non-linear power spectrum. Furthermore we assumed that the NFW profile is unaffected by the dark matter particle mass, whereas at the very least we expect the central core to be smoothed out. We assume that the effect of this on cosmic shear power spectra is subdominant to the decrease in the number of small halos. To overcome this limitation we would need to use a replacement for the NFW which gives the density profile as a function of WDM particle mass and halo mass. In order to truly account for the effects of WDM on halo substructure, one must run N-body simulations (see for e.g. \cite{Colin:2000dn}, \cite{Alam:2002sn} and \cite{Zentner:2003yd}). 

However, our calculations are optimistic in that we use mass functions and mass profiles obtained from simulations containing no baryons. To perform this measurement on real data it would be necessary to use results from simulations using baryons. These are in only the preliminary stages of development at the present time, and there is considerable uncertainty on the prescriptions used as well as on many free parameters that need to enter the calculations. The effect of baryons is to change the predicted cosmic shear power spectra, and to introduce uncertainty in the predictions. We assume for the purpose of this paper that these problems will be solved by the time we will be analysing a Euclid-like survey. We hope that studies like these will provide extra motivation to improve the quality of simulations using baryons. We also assume spherical collapse, which is less realistic in describing non-linear structure than ellipsoidal collapse, which has additional halo mass dependence in the collapse threshold. In other words, a further source of suppression of small scales is the decreased amount of collapse for small haloes, which tend to have higher ellipticity.

Additionally, we normalise the halo bias in order to recover the linear matter power spectrum on very large scales. In fact, the correct approach would be to add a third term to the halo model, which would account for the uncollapsed matter. It would be necessary to specify the power spectrum and correlation of this uncollapsed matter, but it is not clear how best to determine these. We have chosen to normalise the mass-function weighted bias, following convention in CDM halo model calculations. However, the effect of WDM is to decrease the number of low mass halos and therefore the bias is increased on all scales to meet the normalisation condition. This causes an increase in power on mid-range scales.
This may be the reason why the correlations between $\Omega_{\rm m}$ and $m_{\rm WDM}^{-1}$ are inconsistent between the halo model and halofit (See Fig.~\ref{fig:ellipses}). In order to correct this, further studies of WDM on halo bias should be conducted. These issues however only affect the halo model calculations and do not influence our final constraint from the halofit method. A way to include the above suggested correction has been developed by \cite{Smith:2011pr}, where a term is added to the calculation to account for the contribution of uncollapsed matter, which solves the problem of bias normalisation.\\

It is well known that Fisher matrices are an approximate method for forecasting uncertainties on parameters. The formal derivation of the Fisher matrix states that it gives a lower limit on the parameter errors. If the predicted data points are non-linearly related to the parameters, then we can expect likelihood contours in parameter space to be banana shaped, which is not captured within the Fisher matrix errors. Even when considering a single parameter the Fisher matrix approximation breaks down due to the flatness of the log-likelihood, 
giving a gradient of zero at the fiducial value of $m^{-1}_{\rm WDM}=0$. %K
We find this is a significant problem for the WDM particle mass. Small changes in WDM particle mass around our fidicual CDM model make negligible changes in the cosmic shear power spectra which makes it impossible to sensibly use a derivative taken with small step sizes in the Fisher matrix. 
We use CDM as our fiducial model in order to be able to compare our result to other methods (Ly$\alpha$, X-ray). We solve the above problem %K
by using a larger step size, but the Fisher matrix method would ideally be replaced by a full Markov-Chain Monte Carlo exploration of parameter space.

We have assumed that the CMB is insensitive to WDM when we calculated joint constraints with Planck, because the effects of WDM are only important on smaller scales. We have further assumed that only primary CMB anisotropies from Planck are used for constraining cosmology. However, gravitational lensing of the CMB itself can be used to reconstruct single-bin lensing tomography information on the lensing potential (e.g. see \cite{Lewis:2006fu}). This would enhance the constraints on WDM further.

Our cosmic shear calculations rest on a linear theory WDM power spectrum fitting formula which assumes dark matter particles are light and thermalised, like for example a light gravitino. This fitting formula can be straight-forwardly re-expressed for sterile neutrinos and our lower limit on the thermalised WDM particle can be converted to a lower bound on the sterile neutrino mass. For our final result using halofit plus Planck forecasts the limit is $m_{\rm WDM}>2.5$ keV which can be translated to $m_{\nu s}>15.5$ keV. If dark matter is made of sterile neutrinos created using the simplest mechanisms then constraints from cosmic shear can be compared to those from X-ray observations to test the consistency of the model. If cosmic shear were to place the upper limit we derive then it would be inconsistent with current X-ray lower limits and would rule out the simple sterile neutrinos as WDM candidates.

Our forecast constraints from future cosmic shear studies are similar to the upper limits already being obtained from observations of the Lyman-$\alpha$ forest. The main advantage of using cosmic shear is that it does not rely on hydro-dynamical simulations of the intergalactic medium. However the Lyman-$\alpha$ observations are at higher redshift (around $z\sim 3$) than the cosmic shear (which roughly probes mass at $z\lesssim1$). This is advantageous because the matter fluctuations are more in the linear regime and there is less reliance on simulations of non-linear theory. Also the linear theory matter power spectrum is much more sensitive to the WDM particle mass. A deeper cosmic shear survey would gain from this effect.

Cosmic shear power spectra are just one way to probe dark matter structure using gravitational lensing. However, higher order shear statistics such as the bispectrum are more sensitive to small scale clustering than cosmic shear and therefore would be expected to be more powerful in constraining the dark matter mass. Furthermore, higher order distortions of galaxies, such as flexion, also probe smaller scales and would be a better test.
\cite{Bacon:2009aj} find that galaxy-galaxy flexion is sensitive to structure of around $10^{9}$ solar masses. The methodology described in this paper could be extended to make predictions for these statistics.

\acknowledgments
We are grateful to Jesus Zavala for kindly sharing the results of his simulation and to Robert Smith, Adam Amara, Tom Kitching, Alexandre Refregier, Donnacha Kirk, Uro\v{s} Seljak, Alan Heavens, Filipe Abdalla, Martin Kilbinger, Rob Yates, Marco Baldi and John Peacock for many fruitful discussions.
We thank Lawrence Berkeley National Lab \& George Smoot, University College London, CEA Saclay.
KM performed this work with support from the International Max-Planck Research School.
SLB acknowledges support from the Royal Society from a University Research Fellowship, and from an ERC Starting Grant.
AS is supported in part by the U.S. Department of Energy under Contract No. DE-AC02-98CH10886.
We thank the Euclid Weak Lensing Working Group for helpful discussions.

\bibliographystyle{plainnat}
\bibliography{bibliography_2010}

%%%%%%%%%%%%%%%%%%%%%%%%%%%%%%%%%%%%%%%%%%%%%%%
%%%%%%%%%%%%%%%%%%%%%%%%%%%%%%%%%%%%%%%%%%%%%%%
%%%%%%%%%%%%%%%%%%%%%%%%%%%%%%%%%%%%%%%%%%%%%%%
\newpage
\appendix

\section{Appendix: Non-linear power spectrum with the halo model}
\label{sec:appA}

We use the Sheth-Tormen (\cite{Sheth:1999mn}) mass function (see also \cite{Takada:2002qq}, \cite{Seljak:2000gq}):
\begin{equation} \label{eq:massfns2}
n(M)dM=\frac{\rho_{\rm m,0}}{M} A\left[1+(a\nu)^{-p}\right]\sqrt{(a\nu)}e^{-a\nu/2}\frac{d\nu}{\nu} \hspace{1cm} {\rm with} \hspace{1cm} \nu=\left(\frac{\delta_c(z)}{D^+(z)\sigma(M)}\right)^2,
\end{equation}
where $A=0.353$ as in \cite{Jenkins:2000bv}. Also, $a=0.73$ and $p=0.175$. $D^+(z)$ above is the linear growth factor and $\sigma(M)$ is root-mean-square fluctuation in a sphere of volume containing a mass M in a smooth universe, calculated from the linear power spectrum. $\delta_{c}$ is the critical threshold for a linear density fluctuation to collapse to form a halo (see equation \ref{eq:deltac}), calculated from spherical collapse.

From \cite{Takada:2002qq} the bias consistent with the Sheth-Tormen mass functions:
\begin{equation}
b(M)=1+\frac{a\nu-1}{\delta_{\rm c}} +\frac{2p}{\delta_{\rm c}\left(1+(a\nu)^p\right)} \hspace{1cm} {\rm with} \hspace{1cm} \nu=\left(\frac{\delta_{\rm c}(z)}{D^+(z)\sigma(M)}\right)^2   ,
\end{equation}
which must be normalised according to \cite{Seljak:2000gq}.

The non-linear mass scale, $M_{*}(z)$ is defined to be the maximum mass that is undergoing non-linear collapse at a given redshift, which is given by the condition:
\begin{equation} \label{eq:mascale}
\nu(M_*) = \left(\frac{\delta_{\rm c}(z)}{D^+(z)\sigma(M_*)}\right)^2 = 1\hspace{1cm}.
\end{equation}

As mentioned above, we require the threshold for the linear density perturbations to collapse and form a halo. This was given by \cite{Henry:2000bt}:
\begin{equation} \label{eq:deltac}
\delta_c(\Omega_{\rm m,0},z)=\frac{3(12\pi)^{2/3}}{20}\left[1-0.0123\log(1+x^3)\right] \hspace{1cm} {\rm with} \hspace{1cm} x\equiv \frac{\left(\Omega_{\rm m,0}^{-1}-1\right)^{1/3}}{1+z}\hspace{1cm}.
\end{equation}
The density contrast of collapsed objects to the background mass density is taken to be $\Delta(\Omega_{\rm m,0},z)=180$. Therefore the average halo density $\bar{\rho}_{\rm h}(z) = \Delta(z)\rho_{\rm m}(z)$.

The halo density profile is taken to be the Navarro-Frenk-White (NFW) profile with $r_{\rm s}$ being the scale radius defined as the ratio of the virial radius and the halo concentration parameter: $r_{\rm s}=\frac{r_{\rm v}}{c}$,
\begin{equation} \label{eq:nfw}
\rho(r)=\frac{\rho_{\rm s}}{\left(\frac{r}{r_{\rm s}}\right)\left(1+\frac{r}{r_{\rm s}}\right)^2}
\end{equation}
with
\begin{equation}\label{eq:rho_s}
\rho_{\rm s}(M,z) = \frac{180\rho_{\rm m}(z)}{3}\frac{c(M,z)^3}{\left[\ln(1+c(M,z))-\frac{c(M,z)}{1+c(M,z)}\right]} \hspace{1cm}
\end{equation}
and halo mass is defined as
\begin{equation}\label{eq:halomass}
M = \frac{4\pi r_{\rm v}^3}{3} 180\rho_{\rm m}(z) \hspace{1cm} .
\end{equation}
The NFW profile, like other ingredients of the  halo model is calculated from $\Lambda$CDM models. Other halo profiles have been explored by e.g. \cite{Alam:2002sn}.
The halo profile is a function of the halo concentration parameter, c(M,z) as in \cite{Cooray:2000zy}:
\begin{equation} \label{eq:concentration}
c(M,z)=a(z)\left(\frac{M}{M_*(z)}\right)^{b(z)} \hspace{1cm},
\end{equation}
where $a(z) = 10.3(1+z)^{-0.3}$ and $b(z)=0.24(1+z)^{-0.3}$.

The resulting dimensionless power spectrum from \cite{Smith:2002dz} (but also \cite{Cooray:2000ry}, \cite{Seljak:2000gq} etc.), is given by:
\begin{eqnarray}
P^{\rm 1h}_{\rm nl}(k,z)&=&\frac{1}{(2\pi)^3}\int dM\frac{dn}{dM}\left[\frac{\tilde{\rho}(k,M,z)}{\rho_{\rm m,0}}\right]^2\\
P^{\rm 2h}_{\rm nl}(k,z)&=&P_{\rm lin}(k,z)\left[\int dM\frac{dn}{dM}b(M,z)\frac{\tilde{\rho}(k,M,z)}{\rho_{\rm m,0}} \right]^2 \hspace{1cm} , \label{eq:3DmpsH}
\end{eqnarray}
where $\tilde{\rho}(k,M,z)$ is the 3D Fourier transform of the NFW halo density profile. Note that to get the full non-linear power spectrum we must add the 1-halo and 2-halo contributions: $P_{\rm nl}(k,z)=P^{\rm 1h}_{\rm nl}(k,z)+P^{\rm 2h}_{\rm nl}(k,z)$. This is plotted in the right panel of Fig.~\ref{fig:mpswdm}.

\section{Appendix: Step sizes in the Fisher matrix for the inverse WDM particle mass}
\label{appFish}

We first perform a simple likelihood analysis, only varying $m_{\rm WDM}^{-1}$, assuming all other parameters are fixed. We use this to illustrate our proposed method and assess how well it works in one dimension, where we can compare with the truth.

\begin{figure}
\centering
\subfloat{\includegraphics[width=0.5\textwidth]{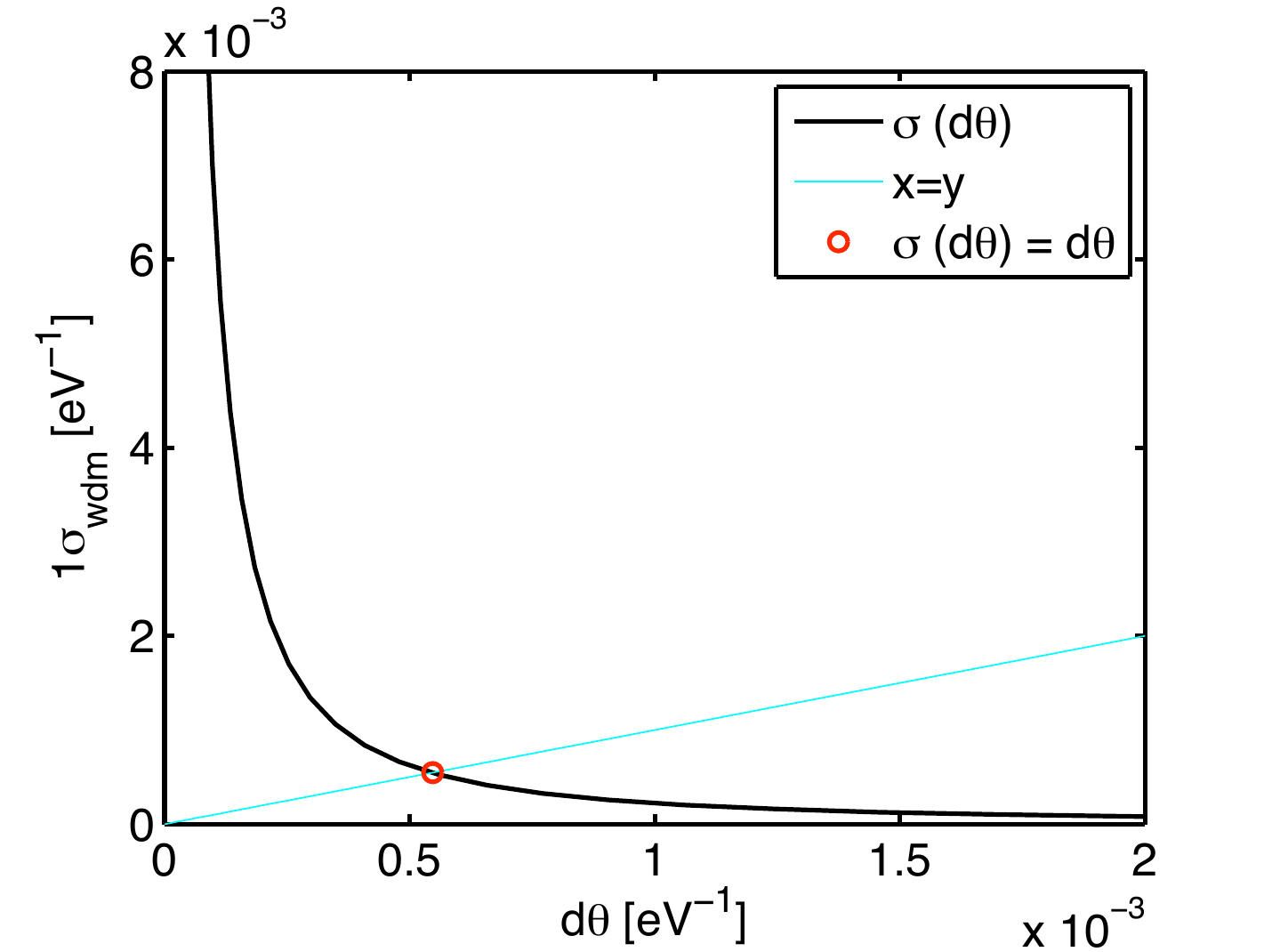}}
\subfloat{\includegraphics[width=0.5\textwidth]{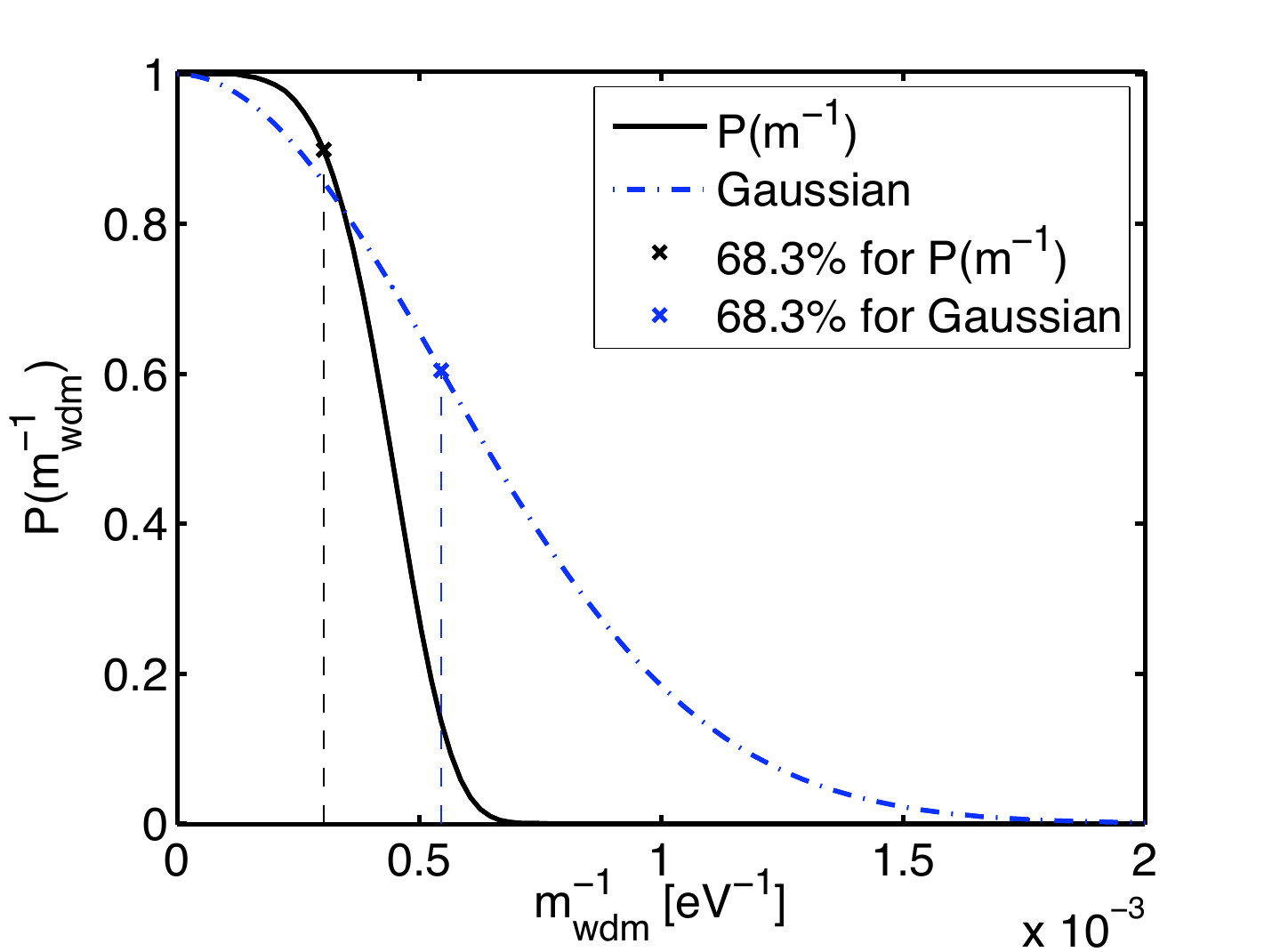}}
\caption{\small
Left: The $1\sigma$ error calculated from Fisher matrices (solid black line), but with different differentiation step sizes used to calculate the gradient of the cosmic shear power spectra with respect to the parameter $m^{-1}_{\rm WDM}$.
The red circle marks the point when the differentiation step size equals the 1$\sigma$ error on the inverse DM particle mass. We fix all other parameters. 
Right: The 1-dimensional probability $P(m_{\rm WDM}^{-1})$ is plotted with a solid black curve. The dot-dashed (blue) curve is the Gaussian probability distribution corresponding to the Fisher matrix error obtained when using a step size equal to the 1$\sigma$ error. The vertical dashed lines correspond to the 68.3 \% of the probability. The red circle on the left corresponds to the vertical dashed blue line on the left.}
\label{fig:chisq}
\end{figure}

As discussed in the main text, we choose to use the inverse WDM particle mass as the free parameter in our Fisher analysis because it has a finite value for our fiducial model, CDM. However as $m^{-1}_{\rm WDM}$ tends to zero the lensing power spectra become completely insensitive to the exact value, which makes it impossible to perform a traditional Fisher matrix analysis in which the step sizes used in calculating the derivatives are very small.
This is illustrated in the left hand panel of Fig.~\ref{fig:chisq} for the simple case where all other parameters are kept fixed. Clearly the choice of step size has a big effect on the result. We propose to use a step size such that it is equal to the error bar, as illustrated by the intersection of the dark (black) and light (cyan) lines in the left hand panel of the figure, shown by the (red) circle. Due to the convex nature of the function this cannot be done by iteration but instead is most simply done by making the figure we show. This can equally well be carried out in multi-dimensional space as in one dimension by working with the one-dimensional marginalised error bar.

By working briefly in one dimension we can compare our result to the truth obtained by plotting the full likelihood as a function of  $m^{-1}_{\rm WDM}$. This is shown in the right hand panel of Fig.~\ref{fig:chisq}. The solid line shows the exact likelihood and the dot-dashed line illustrates the Fisher matrix error obtained by setting the step size equal to the 1$\sigma$ error bar. The crosses and vertical lines show the 68\% one-tailed upper limits from each curve. We see that the exact 68\% upper limit is within a factor of two of the Fisher matrix limit using our proposal. Our proposal gives a more conservative result. We can also see from the difference in shape betwen the solid and dot-dashed lines quite how non-Gaussian the exact probability distribution is. We proceed by using this method in multi-dimensional space.

\end{document}